\documentclass{article} % For LaTeX2e
\usepackage{iclr2026_conference,times}

\usepackage{amsmath,amsfonts,bm}

\def\eqref#1{equation~\ref{#1}}
\def\1{\bm{1}}

\DeclareMathAlphabet{\mathsfit}{\encodingdefault}{\sfdefault}{m}{sl}
\SetMathAlphabet{\mathsfit}{bold}{\encodingdefault}{\sfdefault}{bx}{n}

\usepackage{hyperref}
\usepackage{url}
\usepackage{subcaption}
\usepackage{booktabs}
\usepackage{multirow}
\usepackage{tabularx,array}
\usepackage{amsmath,amsfonts,amsthm}
\usepackage{microtype}
\usepackage[dvipsnames]{xcolor}
\usepackage{todonotes}

\newtheorem{theorem}{Theorem}
\newtheorem{corollary}{Corollary}
 
\title{Training for Compositional Sensitivity Reduces Dense Retrieval Generalization}

\author{Radoslav Ralev, Aditeya Baral, Iliya Zhechev, Jen Agarwal \& Srijith Rajamohan \\
Redis, Bulgaria and Redis, USA\\
\texttt{\{firstname.lastname\}@redis.com} \\
}

\iclrfinalcopy % Uncomment for camera-ready version, but NOT for submission.

\begin{document}
\maketitle

\begin{abstract}
Dense retrieval compresses texts into single embeddings ranked by cosine similarity. While efficient for recall, this interface is brittle for identity-level matching: minimal compositional edits (negation, role swaps) flip meaning yet retain high similarity. Motivated by geometric results for unit-sphere cosine spaces \citep{kang_is_2025}, we test this retrieval-composition tension in text-only retrieval. Across four dual-encoder backbones, adding structure-targeted negatives consistently \emph{reduces} zero-shot NanoBEIR retrieval (8--9\% mean nDCG@10 drop on small backbones; up to 40\% on medium ones), while only partially improving pooled-space separation. Treating pooled cosine as a recall interface, we then benchmark verifiers scoring token--token cosine maps. MaxSim (late interaction) excels at reranking but fails to reject structural near-misses, whereas a small Transformer over similarity maps reliably separates near-misses under end-to-end training. \footnote{Code and datasets are available at \url{https://github.com/radoslavralev/limitations-text-retrieval}}
\end{abstract}

\section{Introduction}
\label{sec:introduction}

The dominant dual-encoder paradigm compresses texts into fixed vectors for efficient maximum inner product search (MIPS) retrieval \citep{reimers-gurevych-2019-sentence, karpukhin-etal-2020-dense}. While effective for fuzzy topical matching, this architecture suffers a fundamental ``resolution loss'' regarding composition. Because the embedding function compresses variable-length reasoning into a single point, it often treats sentences as commutative bags-of-words, struggling to distinguish \emph{structural near-misses} (e.g.,``the dog bit the man'' vs.\ ``the man bit the dog'') \citep{yuksekgonul_when_2022}.

Recent theory suggests this is geometrically inevitable: \citet{kang_is_2025} argue that unit-sphere cosine spaces force conceptual clusters into linear superposition, a geometry hostile to non-commutative structures like negation or order. This implies a \emph{retrieval--composition tension}: forcing compositional sensitivity into a single vector degrades broad topical generalization.

\textbf{Contributions.} We investigate this tension in text-only retrieval. We show that training with structure-targeted hard negatives creates a zero-sum game: the model rejects specific permutations but suffers significant degradation in out-of-domain retrieval (NanoBEIR). We argue that identity-sensitive matching should instead be treated as a distinct \emph{verification} task. We benchmark lightweight verifiers on token--token similarity maps, finding that while MaxSim excels at relevance, true identity preservation requires learned verifiers that detect topological patterns in the map.

\section{Single‑vector cosine is a bottleneck for identity}
Under unit-norm pooled embeddings and cosine scoring, a single inner product must simultaneously encode topical similarity and compositional distinctions.
Previous work asserts that nontrivial content grouping pressures the representation toward (approximately) additive superposition \citep{kang_is_2025}, which is commutative and tends to erase binding/order information.
This predicts brittleness: there exist minimally edited near-misses (binding swaps, role reversals, scoped negation flips) that cannot be uniformly separated from paraphrases by a fixed cosine margin under the pooled-cosine bottleneck.
We include the formal assumptions and an expanded statement in Appendix~\ref{app:theory}.

We adopt the standard two-stage setup.
\textbf{Stage 1:} retrieve top-$K$ candidates using ANN over pooled cosine keys.
\textbf{Stage 2:} verify candidates using token interactions.

Given token embeddings for query $q$ and candidate $c$, we form the token similarity map
$M_{ij}(q,c)=\cos(q_i,c_j)$.
A verifier $F(q,c)$ consumes $M$ (optionally with positional bias) and outputs a scalar used to rerank or gate candidates.
We study a spectrum from simple reductions (global average; MaxSim/late interaction) to small learned pattern recognizers over $M$ (tiny CNN / tiny Transformer).
Full definitions (including alignment-biased variants and architectures) are in Appendix~\ref{app:verifiers}.

\section{Experiments}
\label{sec:experiments}

Our analysis predicts a \emph{retrieval--composition tension} for pooled-cosine dual encoders: allocating representational margin to reject meaning-changing near-misses can reduce the margin available for coarse content grouping.
We test:
(i) whether structure-targeted hard negatives degrade out-of-domain retrieval, and
(ii) what verifier capacity is required to reject structural near-misses. For more information on dataset generation see Appendix \ref{app:dataset-construction}.

\subsection{Do composition-sensitive negatives hurt retrieval?}
\label{subsec:filtering}

We fine-tune dual encoders on \textsc{NQ} triplets using SentenceTransformers' MultipleNegativesRankingLoss.
We compare:
\textbf{Model A (baseline)} trained on standard \textsc{NQ} supervision, and
\textbf{Model B (structured)} trained on the mixed dataset described in \S\ref{app:dataset-construction} (standard + structural negatives).
To compare across backbones under a fixed compute budget, we fix wall-clock training time per backbone and set steps based on measured throughput (details in Appendix). We evaluate zero-shot retrieval on NanoBEIR using nDCG@10 and Acc@1 (mean across datasets).
Table~\ref{tab:nanobeir-mean-summary} summarizes mean results across four backbones.

\begin{table}[t]
\centering
\caption{Mean NanoBEIR retrieval performance (nDCG@10 and Acc@1). Model A: standard fine-tuning. Model B: + structured negatives.}
\resizebox{\linewidth}{!}{%
\begin{tabular}{l ccc ccc}
\toprule
 & \multicolumn{3}{c}{\textbf{nDCG@10}} & \multicolumn{3}{c}{\textbf{Acc@1}} \\
\cmidrule(lr){2-4} \cmidrule(lr){5-7}
\textbf{Backbone} & Model A & Model B & $\Delta$ (\% drop) & Model A & Model B & $\Delta$ (\% drop) \\
\midrule
MiniLM-L6 & 0.439$\pm$0.000 & 0.401$\pm$0.001 & \textbf{-0.038 (-8.7\%)} & 0.393$\pm$0.002 & 0.346$\pm$0.004 & \textbf{-0.047 (-12.0\%)} \\
MiniLM-L12 & 0.467$\pm$0.001 & 0.424$\pm$0.005 & \textbf{-0.043 (-9.2\%)} & 0.424$\pm$0.003 & 0.369$\pm$0.010 & \textbf{-0.055 (-13.0\%)} \\
GTE-Small & 0.481$\pm$0.002 & 0.442$\pm$0.006 & \textbf{-0.039 (-8.1\%)} & 0.444$\pm$0.001 & 0.389$\pm$0.004 & \textbf{-0.055 (-12.4\%)} \\
GTE-ModernBERT-base & 0.543$\pm$0.001 & 0.324$\pm$0.018 & \textbf{-0.219 (-40.3\%)} & 0.493$\pm$0.005 & 0.275$\pm$0.015 & \textbf{-0.218 (-44.2\%)} \\
\bottomrule
\end{tabular}
}
\label{tab:nanobeir-mean-summary}
\end{table}

\paragraph{Results.}
Across all backbones and metrics, training with structural hard negatives (\textbf{Model B}) reduces NanoBEIR performance relative to the \textsc{NQ}-only baseline (\textbf{Model A}).
On MiniLM-L6/L12 and \texttt{gte-small}, mean nDCG@10 drops by 8--9\% and Acc@1 drops by 12--13\%.
On \texttt{gte-modernbert-base}, the drop is much larger (40\% nDCG@10; 44\% Acc@1).
This supports the predicted tension: under a single pooled embedding with cosine scoring, allocating margin to reject lexically overlapping meaning-changes competes with broad topical grouping.

\begin{figure}[t]
    \centering
    \includegraphics[width=0.8\linewidth]{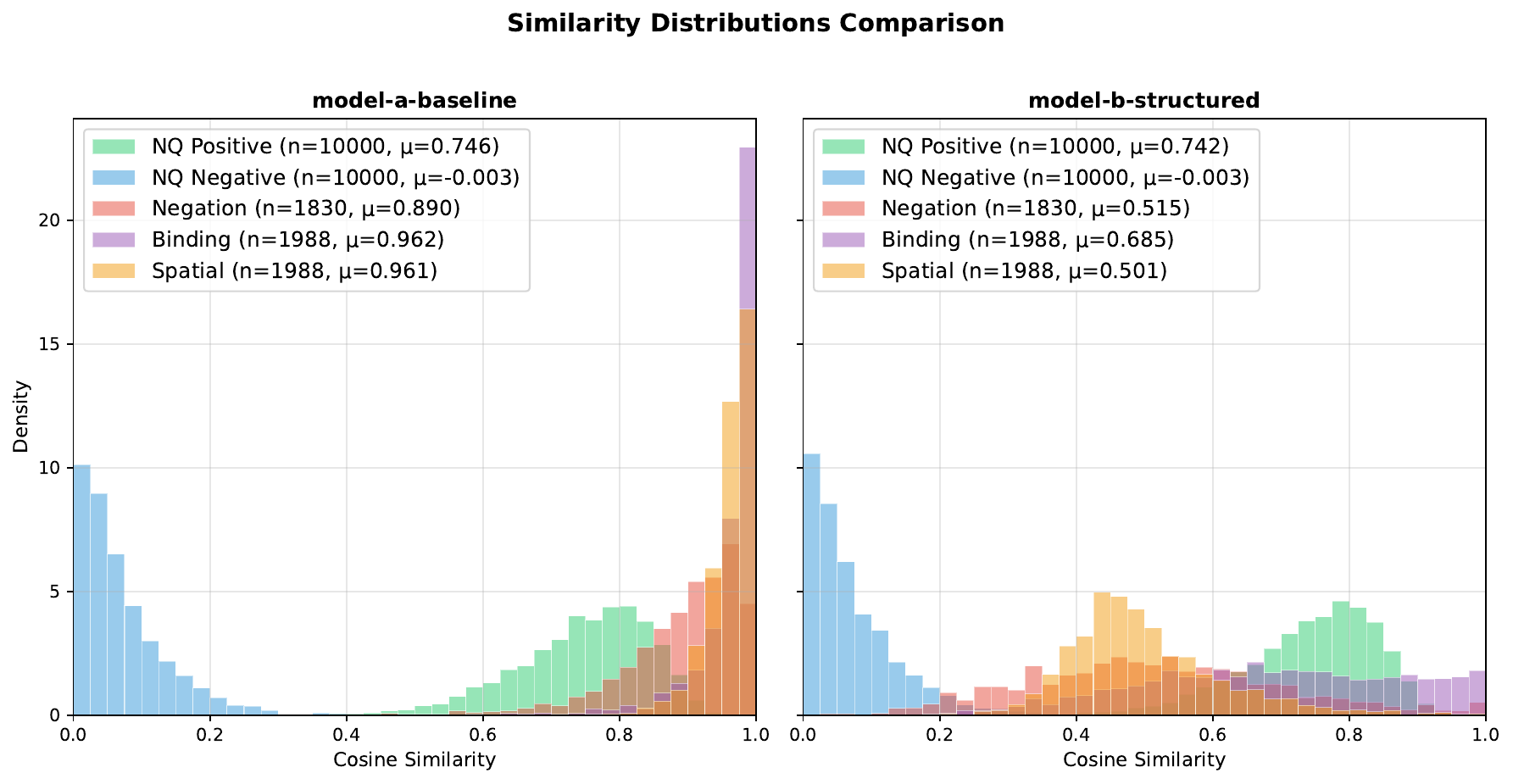}
    \caption{Cosine-similarity distributions between an anchor sentence and a minimally edited near-miss under pooled embeddings.
We compare Model~A vs.\ Model~B for three perturbation families (negation, binding/order, spatial) and overlay \textsc{NQ} positives/negatives for reference (10k pairs each).
Lower is better for near-miss distributions.}

    \label{fig:cosine_distributions}
\end{figure}

\paragraph{Does the retrieval drop buy identity sensitivity in pooled space?}
To measure what compositional sensitivity is obtained \emph{within the pooled space}, we plot cosine-similarity distributions between an original sentence $s$ and a minimally perturbed near-miss $\tilde{s}$ (negation, binding/order, spatial flips).
Lower cosine is better: all perturbations are non-identical by construction.
Fig.~\ref{fig:cosine_distributions} overlays these distributions with 10k held-out \textsc{NQ} positives and negatives.

Two patterns stand out.
First, \textsc{NQ}-only fine-tuning (Model~A) leaves identity-breaking edits highly similar to the anchor: negation and binding remain near the positive regime, and spatial flips are nearly saturated.
Second, introducing structural negatives (Model~B) produces \emph{non-uniform} improvements: while it significantly reduces similarity for negation and spatial flips, the gains for binding are less definitive. Despite a lower mean, binding lacks a distinct cluster to separate it from other categories. Thus, while structure-targeted negatives improve sensitivity for specific perturbation classes, they fail to establish a consistent identity margin in pooled cosine space, underscoring the continued necessity of token-interaction verification.

\paragraph{Takeaway:} structural negatives partially lower cosine for some edits but reliably hurt out‑of‑domain retrieval.
\subsection{How small can the verifier be?}
\label{subsec:verifier-scaling}

We evaluate the verifier family $\{F_k\}$ from \S\ref{subsec:verifiers} operating over token--token cosine maps $M(q,c)$.
We compare:
(i) \textbf{Frozen encoder}, where we train only the verifier, and
(ii) \textbf{End-to-end}, where we train encoder and verifier jointly.
All methods share the same stage-1 candidate generation via pooled cosine; only the stage-2 verifier differs.

\begin{figure}[t]
    \centering
    \includegraphics[width=0.9\linewidth]{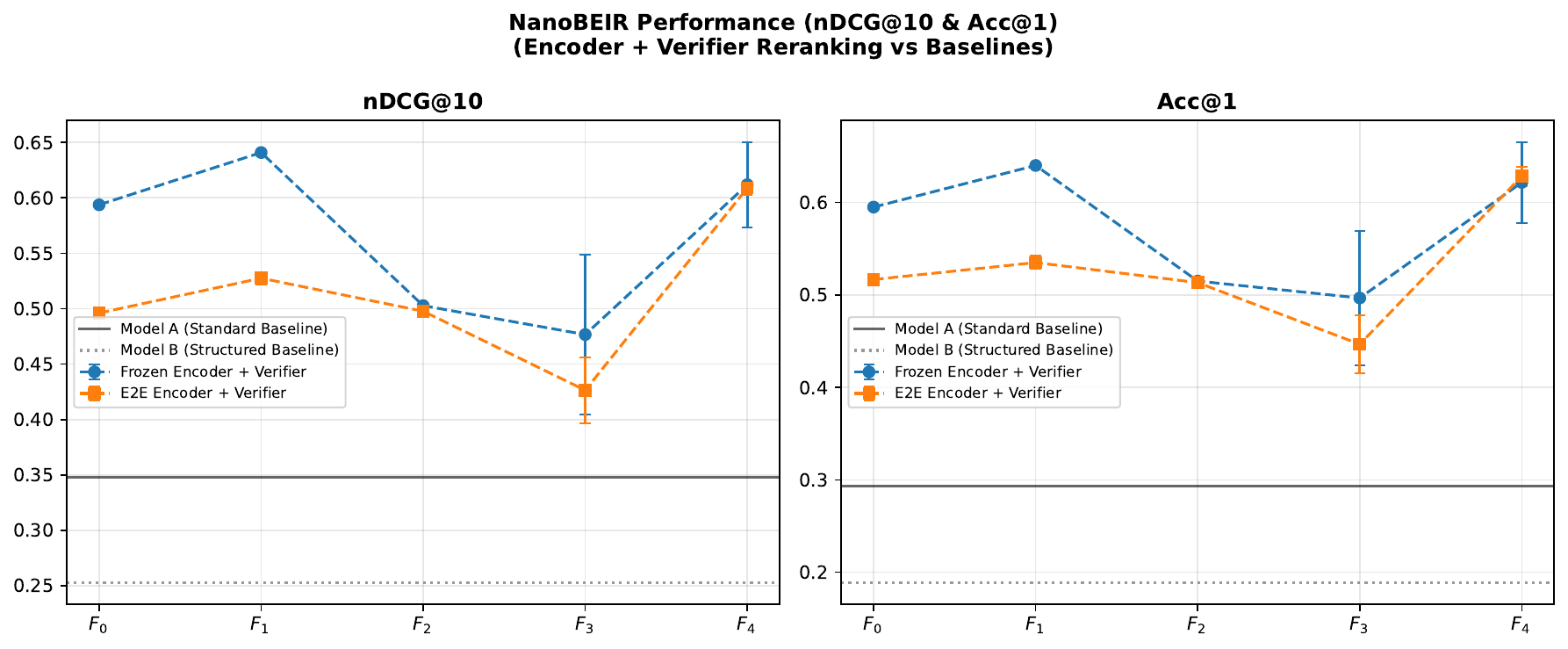}
    \caption{NanoBEIR performance after reranking top-$K$ candidates with $F_k$ under a frozen-encoder (blue) or end-to-end (orange) regime; horizontal lines show encoder-only baselines (Model~A and Model~B). MaxSim ($F_1$) is the strongest frozen reranker; end-to-end $F_4$ is most competitive.}
    \label{fig:nanobeir_verifier}
\end{figure}

\paragraph{Evaluation 1: reranking on NanoBEIR.}
Fig.~\ref{fig:nanobeir_verifier} reports NanoBEIR metrics after reranking the top-$K$ candidates with each verifier.
In the \textbf{frozen} regime, late interaction $F_1$ (MaxSim) is the strongest and most consistent reranker across metrics; $F_0$ and $F_4$ are often close, while soft alignment $F_2$ is consistently weaker.
In the \textbf{end-to-end} regime, verifier choice matters more: jointly training with the map-Transformer $F_4$ yields the largest and most reliable gains.

\begin{figure}[t]
    \centering
    \includegraphics[width=0.9\linewidth]{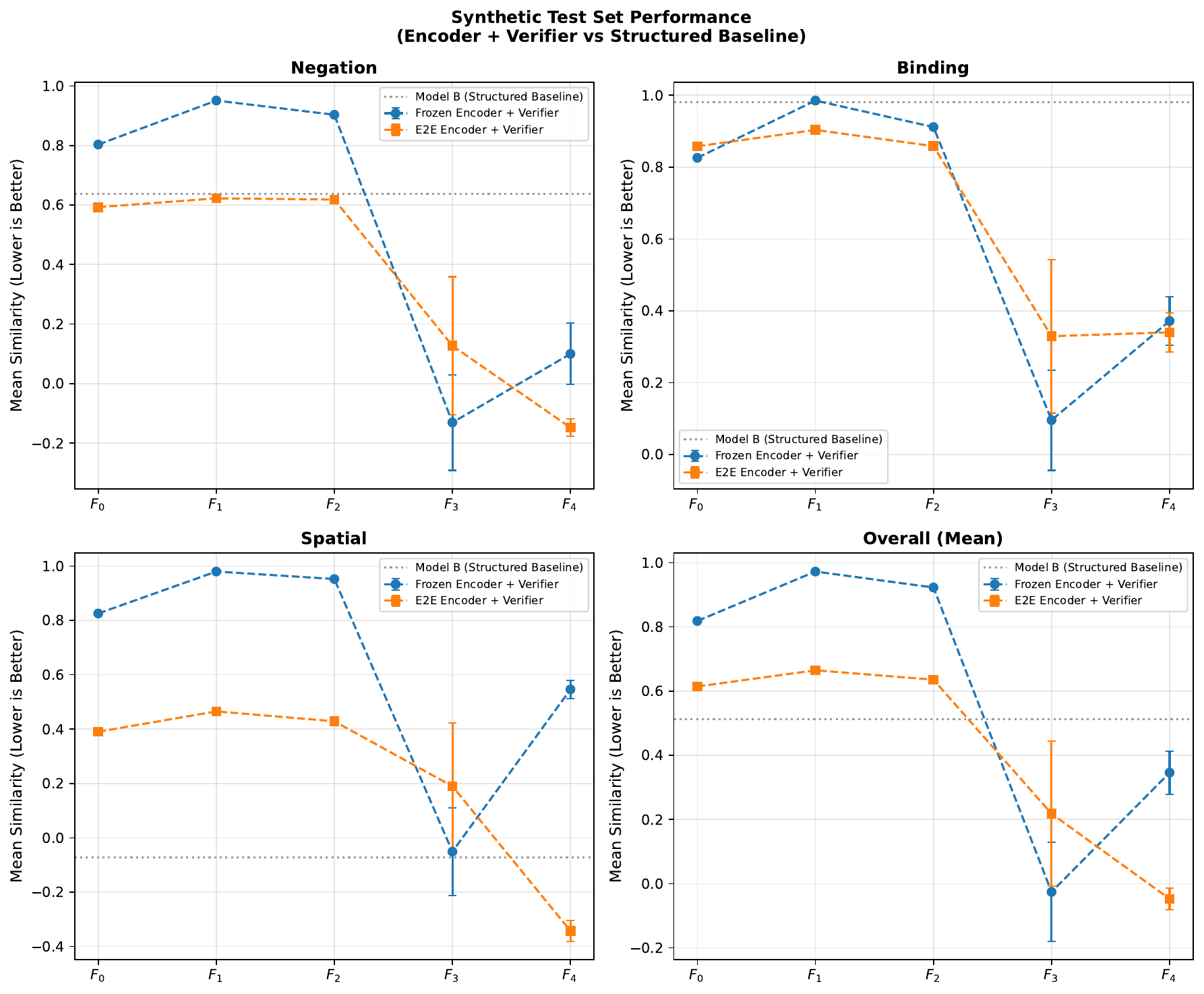}
    \caption{Synthetic structural near-miss test. Mean scores on hard negatives (near-misses); lower is better. The dotted line is pooled cosine from Model~B. Simple reductions of $M$ ($F_0$--$F_2$) and MaxSim ($F_1$) score near-misses as highly similar, while topology-aware verifiers ($F_3$, $F_4$) substantially reduce near-miss scores; end-to-end $F_4$ is strongest on spatial flips.}
    \label{fig:synthetic_verifier}
\end{figure}

\paragraph{Evaluation 2: synthetic structural near-miss test.}
We evaluate on the held-out 5{,}964-pair split from \S\ref{app:dataset-construction}, grouped into \textbf{Negation}, \textbf{Binding/Order}, and \textbf{Spatial}.
Fig.~\ref{fig:synthetic_verifier} plots the mean score assigned to near-miss pairs (lower is better). The dotted horizontal line shows the pooled-cosine score from the structured encoder baseline (Model~B).

\paragraph{Results.}
Comparing Fig.~\ref{fig:nanobeir_verifier} and Fig.~\ref{fig:synthetic_verifier} highlights a key mismatch.
MaxSim ($F_1$) improves benchmark reranking on NanoBEIR but fails to reject structural near-misses, assigning them near-identity scores.
Conversely, learned map-based verifiers ($F_3$/$F_4$) substantially improve near-miss separation, with $F_4$ strongest under end-to-end training, but are not always the top frozen rerankers.
This reinforces that if a deployment requires identity-level correctness, verification must be treated as a distinct objective with appropriate data and calibration, rather than assumed to follow from relevance benchmarks.

\paragraph{Takeaway:} MaxSim is a strong relevance reranker, but identity rejection needs learned map structure.

\section{Discussion and conclusion}
Pooled-cosine embeddings are a strong \emph{recall} interface for content grouping, but our results support a structural limitation for identity-sensitive matching: injecting identity-focused negatives into a single-vector objective can trade off against out-of-domain relevance retrieval.
Token-interaction verification is a principled escape hatch, but relevance reranking (NanoBEIR) and identity rejection are not automatically aligned: MaxSim helps the former while failing the latter, whereas small learned verifiers over similarity maps better enforce compositional identity.
This motivates treating identity-sensitive verification as a distinct objective with dedicated data and calibration.

\section*{Reproducibility Statement}
Complete experimental settings (model architectures, hyperparameters, preprocessing, random seeds, hardware/software versions, and evaluation protocol) are provided in Appendix~\ref{app:expdetails}.
The shared anonymized repository includes the code used to train and evaluate all models, scripts for dataset construction, and the exact dataset splits used in our experiments.

\section*{Ethics Statement}
We adhere to the ICLR Code of Ethics. Our experiments use only publicly available benchmark datasets and automatically constructed structural near-miss examples; we collect no new user data and involve no human subjects. We comply with dataset licenses and will release only license-compliant artifacts. Potential risks include biased retrieval/verification behavior inherited from pretrained models or dataset distributions; we recommend auditing before deployment in sensitive applications.

\bibliographystyle{iclr2026_conference}
\bibliography{iclr2026_conference}

% =========================
% APPENDIX (unlimited pages)
% =========================
\appendix

\section{Expanded related work}
\label{app:related}
% Moved from main: Related Work section.
\noindent\textbf{Pooled embeddings and compositional failures.}
Single-vector cosine embeddings enable fast ANN retrieval but often under-encode binding, order, and scoped negation; stress tests find strong retrieval despite compositional ablations, suggesting shortcut solutions \citep{yuksekgonul_when_2022,kamath_whats_2023,hsieh_sugarcrepe_2023,alhamoud_vision-language_2025}.

\noindent\textbf{Geometric analyses and token-interaction remedies.}
\citet{kang_is_2025} show that cosine spaces satisfying basic categorization induce linear superposition, collapsing attribute binding and conflicting with spatial relations and negation; they propose Dense Cosine Similarity Maps and lightweight CNNs over interactions.

\noindent\textbf{Two-stage retrieval and verification.}
Candidate generation plus reranking is standard: cross-encoders compute full interactions, while late interaction retains token structure with the efficient MaxSim operator \citep{nogueira_passage_2019,nogueira_document_2020,khattab_colbert_2020,santhanam_colbertv2_2022}.
Sparse expansions (e.g., SPLADE) offer an alternative first-stage representation \citep{formal_splade_2021}.

\noindent\textbf{Indexing and compression.}
ANN systems and quantization are standard for dense retrieval \citep{johnson_billion-scale_2018,malkov_efficient_2018,jegou_product_2011,ge_optimized_nodate}.

\noindent\textbf{Embedding geometry.}
Work on anisotropy and cosine similarity supports structured scoring beyond pooled cosine \citep{ethayarajh_how_2019,steck_is_2024}.

\section{Theory details: pooled-cosine brittleness}
\label{app:theory}
% Moved from main: Background + Theorem/Corollary discussion.
% Keep your original text here, potentially verbatim, including:
Many semantic search deployments are \emph{content-relevance} oriented regardless of fine-grained semantic differences. However, several important applications require \emph{identity-sensitive} matching: the system must accept a candidate only if it expresses the same proposition up to paraphrase, rejecting candidates with nearly identical wording but different meaning or intent (see examples in \S\ref{sec:introduction}).
We treat as \emph{non-identical} (near-miss negatives) edits that change:
(i) \emph{attribute--head binding} (which modifier applies to which head),
(ii) \emph{relations and argument roles/order} (subject/object swaps, attachment changes),
or (iii) \emph{negation and scope} (polarity flips or changes in what an operator negates).

\subsection{Single-vector cosine retrieval}
Let $\mathcal{V}$ be a vocabulary and $\mathcal{S}\subseteq \mathcal{V}^{\ast}$ the set of well-formed sentences (or clauses).
We study \emph{text-only} embedding-based semantic search systems that map each $s\in\mathcal{S}$ to a single vector and use ANN search to retrieve candidates. We write $q \equiv c$ when $q$ and $c$ express the same proposition.

Let $e_{\theta}:\mathcal{S}\rightarrow\mathbb{S}^{d-1}$ map each sentence to a \emph{unit} vector in $\mathbb{R}^d$.\footnote{We write $\mathbb{S}^{d-1}=\{u\in\mathbb{R}^d:\|u\|_2=1\}$.}
A standard match surrogate is cosine thresholding,
\begin{equation}
\textsf{accept}_{\tau}(q,c)\;=\;\mathbf{1}\!\left[\cos\!\big(e_{\theta}(q),e_{\theta}(c)\big)\ge\tau\right].
\label{eq:accept}
\end{equation}
This interface enables compact indexes and efficient ANN search, but it enforces a severe bottleneck: all semantics must be encoded into a single direction on the sphere, and the decision depends on a single inner product.

\subsection{Why pooled cosine is brittle for compositional identity}
\label{subsec:ideal-geometry-text}

Our analysis follows the \emph{ideal-geometry} framework of \citet{kang_is_2025}.
They formalize conditions for an ``ideal'' CLIP-like unit-sphere cosine space and prove these conditions are mutually incompatible: satisfying basic concept categorization forces a linear superposition geometry that cannot also satisfy binding, spatial relations, and negation.
We adapt the implication to text-only retrieval; full formal definitions and proofs are in \citet{kang_is_2025} (and its supplement), and we focus primarily on empirical consequences for text retrieval.

\textbf{Content grouping and superposition.}
Dense retrievers are typically trained/evaluated so that texts sharing salient content words or topics are closer than texts with disjoint content.
Under unit-norm embeddings with cosine scoring, \citet{kang_is_2025} show that the cosine-optimal representation for a composition that must remain close to its constituents is (approximately) a normalized linear superposition.
In text terms, if a sentence expresses salient units $x_1,x_2\in\mathcal{V}$ and must remain close to each while repelling unrelated content, then
\begin{equation}
e_{\theta}(x_1\,x_2)\;\approx\;\frac{e_{\theta}(x_1)+e_{\theta}(x_2)}{\|e_{\theta}(x_1)+e_{\theta}(x_2)\|}.
\label{eq:superposition-text}
\end{equation}
Superposition is commutative; without additional structure at scoring time, it naturally encourages invariances that erase binding and role information.

\textbf{Minimal identity constraints.}
For identity-sensitive matching, we would like paraphrases $q^{+}\equiv q$ to be closer than minimally edited near-misses $q^{-}\not\equiv q$ by a margin:
\begin{equation}
\cos(e_{\theta}(q),e_{\theta}(q^{+})) \;\ge\; \cos(e_{\theta}(q),e_{\theta}(q^{-})) + \gamma.
\label{eq:margin}
\end{equation}
Near-misses include (i) binding swaps, (ii) role/order reversals, and (iii) negation/scope flips.

\textbf{Assumptions.}
We isolate the interface shared by most embedding retrievers:
\begin{description}
\item[A1] \emph{Single pooled key:} each sentence is represented by one unit vector in $\mathbb{S}^{d-1}$.
\item[A2] \emph{Cosine scoring:} decisions depend only on cosine similarity between pooled keys.
\item[A3] \emph{No token interactions at score time:} the scorer has no access to token--token alignments beyond what is compressed into the pooled key.
\end{description}

\begin{theorem}[Informal pooled-cosine brittleness for compositional identity]
\label{thm:text-impossibility}
Under A1--A3, any encoder family that enforces nontrivial content grouping (compositions remain close to their constituents with margin) necessarily admits clause pairs that differ only by
(i) attribute binding, (ii) relational roles/order, or (iii) negation/scope,
yet cannot be simultaneously separated from identity-preserving paraphrases by a fixed cosine margin.
\end{theorem}

\noindent\emph{Justification}
Content grouping implies an approximately additive/superpositional placement (Lemma~1 in \citet{kang_is_2025}); commutativity yields binding collapse (Lemma~2) and analogous invariances for role/order.
When one additionally enforces natural cosine behavior for negation, \citet{kang_is_2025} derive further contradictions.
We omit the full formalization for text and refer to \citet{kang_is_2025} for complete proofs.

\begin{corollary}[Threshold brittleness]
\label{cor:text-brittleness}
If A1--A3 hold and content grouping has margin $\gamma_{\mathrm{cont}}>0$, then for any fixed threshold $\tau$ there exist minimally edited near-miss pairs $(q,c)$ (binding swap, role reversal, or scoped negation flip) such that Eq.~\eqref{eq:accept} incurs either a false accept or a false reject at a scale comparable to $\gamma_{\mathrm{cont}}$.
\end{corollary}

A practical implication is a \emph{retrieval--composition tension}: if we insist on a single pooled key and cosine as the only scoring mechanism, encoding fine-grained structure competes with the angular budget used for coarse content grouping.
In \S\ref{sec:experiments}, we test whether structure-targeted hard negatives produce this trade-off in text-only dual-encoder training.

% (Paste your original Section 3 text here.)

\section{Verifier definitions and architectures}
\label{app:verifiers}
% Moved from main: Full F0--F4 equations and soft alignment.
Theorem~\ref{thm:text-impossibility} points to an interface mismatch: the bottleneck is not necessarily the token representations themselves, but the fact that the final decision collapses everything into one cosine score.
A natural remedy---already prevalent in IR---is a two-stage pipeline: use pooled embeddings for high-recall candidate generation, then \emph{verify} (or rerank) with token-level interactions \citep{nogueira_passage_2019,khattab_colbert_2020}.

\subsection{Two-stage retrieval with token-level verification}
\label{subsec:two-stage}

\textbf{Stage 1 (candidate generation).}
A transformer encoder produces contextual token embeddings
$H_{\theta}(s)=[h_1,\dots,h_{m(s)}]\in\mathbb{R}^{m(s)\times d}$.
We pool to a unit key $e_{\theta}(s)\in\mathbb{S}^{d-1}$ (CLS/mean/EOS) and retrieve top-$K$ candidates with ANN under cosine similarity.

\textbf{Stage 2 (verification).}
For a query $q$ and candidate $c$ with token embeddings
$Q=[q_1,\dots,q_m]$ and $C=[c_1,\dots,c_n]$, define the token similarity map
\begin{equation}
M(q,c)\in[-1,1]^{m\times n},
\qquad
M_{ij}(q,c)=\cos(q_i,c_j).
\label{eq:ttsm}
\end{equation}
Here $\phi$ denotes elementwise normalization/clipping of $M$, and $\psi$ patches (or flattens) the map into a sequence for the Transformer. A verifier consumes $M(q,c)$ (optionally with positional information) and outputs a scalar score $F(q,c)$ used for gating or reranking. 

\subsection{A spectrum of lightweight verifiers}
\label{subsec:verifiers}

We study verifiers $\{F_k\}$ that vary in expressivity/cost while remaining far cheaper than full cross-encoding over long corpora.
All verifiers operate on $M$ after stage-1 retrieval.

\begin{align}
F_0(q,c) &= \tfrac{1}{mn}\sum_{i=1}^{m}\sum_{j=1}^{n} M_{ij}
&& \text{(global average)} \label{eq:F0}\\[2pt]
F_1(q,c) &= \tfrac{1}{m}\sum_{i=1}^{m}\max_{j} M_{ij}
&& \text{(late interaction / MaxSim)}
 \label{eq:F1}\\[2pt]
F_2(q,c) &= \tfrac{1}{m}\sum_{i=1}^{m}\sum_{j=1}^{n} A_{ij}(q,c)\, M_{ij}
&& \text{(soft alignment with positional bias)} \label{eq:F2}\\[2pt]
F_3(q,c) &= \mathrm{MLP}\!\Big(\mathrm{CNN}_{k\times k}\big(\phi(M)\big)\Big)
&& \text{(tiny CNN over $M$)} \label{eq:F3}\\[2pt]
F_4(q,c) &= \mathrm{MLP}\!\Big(\mathrm{Transformer}\big(\psi(\phi(M))\big)_{\mathrm{[CLS]}}\Big)
&& \text{(tiny Transformer over patches of $M$)} \label{eq:F4}
\end{align}
where $A(q,c)$ is a row-stochastic alignment matrix:
\begin{equation}
A_{ij}(q,c)
=\frac{\exp\!\big((M_{ij}(q,c)-\lambda|i-j|)/\tau\big)}{\sum_{k=1}^{n}\exp\!\big((M_{ik}(q,c)-\lambda|i-k|)/\tau\big)}.
\label{eq:F2_soft_align}
\end{equation}

\subsection{Why token interactions help}
\label{subsec:why-works}

The pooled-cosine bottleneck collapses many compositions because it discards token topology.
By contrast, $M(q,c)$ preserves which tokens align and \emph{where} those alignments occur.
Verifiers that only aggregate $M$ with permutation-symmetric statistics (e.g., $F_0$, and to a large extent $F_1$) can still behave like bag-of-words matchers and remain insensitive to binding or role swaps.
Injecting positional structure (as in $F_2$) and learning local/global patterns over $M$ (as in $F_3$/$F_4$) breaks these symmetries, allowing the verifier to detect order-preserving diagonals, swapped alignments, and systematic mismatches induced by negation cues.
This mirrors the core insight of DCSMs in \citet{kang_is_2025}, specialized here to text--text matching.

\section{Experimental details}
\label{app:expdetails}

This section summarizes the datasets, model variants, training setup, and evaluation protocol needed to reproduce our results.

\subsection{Data}
\label{app:dataset-construction}

\paragraph{Baseline training data (Natural Questions).}
We fine-tune dual encoders on 100{,}000 triplets sampled from Natural Questions \citep{natquest} using the standard
\texttt{(anchor, positive, negative)} format.

\paragraph{Structural hard negatives.}
We augment training with \emph{structural near-misses}: lexically high-overlap pairs whose meaning differs due to
(i) negation/scope flips, (ii) binding/order changes, or (iii) spatial relation flips.
We construct 9{,}940 pairs per category (29{,}820 total) and convert each pair $(s_1,s_2)$ into a triplet
$(s_1, s_1, s_2)$ so the model must repel the near-miss while keeping the anchor fixed.
We split pairs 80/20 and use the held-out split (5{,}964 pairs) for synthetic evaluations.

The final structured-training mixture contains 123{,}856 triplets, where 23{,}857 (19.2\%) are structural-negative
triplets and the remainder are standard \textsc{NQ} triplets.
We drop null/placeholder rows, filter sentences shorter than 20 characters, and truncate/pad to 128 tokens.

\subsection{Models}
\label{app:models}

\paragraph{Stage-1 candidate generators (dual encoders).}
We evaluate four backbones:
\texttt{sentence-transformers/all-MiniLM-L6-v2},
\texttt{sentence-transformers/all-MiniLM-L12-v2},
\texttt{thenlper/gte-small}, and
\texttt{Alibaba-NLP/gte-modernbert-base}.
We use the default pooling method of each encoder, max length 128, and unit-normalized pooled embeddings with cosine scoring.
MiniLM and \texttt{gte-small} use 384-d pooled embeddings; other backbones use their native embedding dimensions.

\paragraph{Stage-2 verifiers.}
Verifiers consume token--token cosine maps $M(q,c)$ and output a scalar score for reranking/gating
(Appendix~\ref{app:verifiers}). We evaluate $F_0$--$F_4$ as defined in Appendix~\ref{app:verifiers}.
Learned verifiers use small networks over $M$ (a tiny CNN for $F_3$ and a tiny Transformer for $F_4$).

\subsection{Training}
\label{app:training}

\paragraph{Encoder training objective.}
We fine-tune using SentenceTransformers' MultipleNegativesRankingLoss with temperature $\tau{=}0.1$,
optimized with AdamW and a linear warmup/decay schedule.

\paragraph{Key hyperparameters.}
Unless otherwise stated: learning rate $2{\times}10^{-5}$ (scaled by model size in code), weight decay 0.01,
batch size 64 (and 128 in selected runs), warmup ratio 0.1, gradient accumulation 1, fp16/bf16 precision.
We fix wall-clock training time per backbone and set steps based on measured throughput.

\paragraph{Verifier training.}
We compare (i) \textbf{Frozen} (train verifier only) and (ii) \textbf{End-to-end} (train encoder+verifier jointly).
Verifier LR is $1{\times}10^{-4}$; end-to-end encoder LR is $1{\times}10^{-5}$ (scaled by model size in code).
Batch size is 128 for $F_0$--$F_2$ and 32 for $F_3$--$F_4$. We early-stop with patience 5000 steps on nDCG@10.

\paragraph{Random seeds.}
Primary seed is 42. Multi-seed results use seeds $\{42,43,44\}$.

\subsection{Evaluation protocol}
\label{app:eval}

\paragraph{Retrieval benchmarks.}
We evaluate zero-shot retrieval on NanoBEIR (\texttt{lightonai/NanoBEIR-en}) and report mean performance across datasets.
We report nDCG@10 and Acc@1 in the main paper (additional metrics are computed in code).

\paragraph{Two-stage evaluation.}
Stage 1 retrieves top-$K{=}100$ candidates using pooled-cosine ANN.
Stage 2 (optional) reranks/gates the top-$K$ using a verifier score. Evaluation batch size is 32.

\subsection{Compute and software}
\label{app:compute}

We run on GPUs with $\ge$24GB VRAM (tested on NVIDIA L4 and A10-class hardware).
Typical training time is $\sim$4 minutes per configuration; the full experiment suite runs in $\sim$2--3 hours.
We use Python 3.10 with PyTorch, HuggingFace Transformers, SentenceTransformers, and BEIR; exact versions are pinned in the released environment files.
\end{document}